\documentclass[preprint,showpacs,preprintnumbers,amsmath,amssymb]{revtex4}

\usepackage{graphicx}
\usepackage{dcolumn}
\usepackage{bm}

\begin{document}



\title{Scale-Dependent Price Fluctuations for the Indian
Stock Market}


\author{Kaushik~Matia$^1$, Mukul~Pal$^2$, H. Eugene~Stanley$^1$, H.~Salunkay$^3$
}

\affiliation{
$^1$Center for Polymer Studies and Department of Physics, Boston University, Boston, MA 02215. \\
$^2$BSE Training Institute, The Stock Exchange Mumbai, P J Towers, Mumbai, India.\footnote{Present address: Edelweiss Capital Ltd. 1st Floor Shalaka, Maharshi Karve Marg, Cooperage, Mumbai 400021, India}  \\
$^3$Online Derivatives, 607 Reena Complex, Vidya Vihar (West), Mumbai 400086, India. \\
}




\begin{abstract}

Classic studies of the probability density of price fluctuations $g$ for
stocks and foreign exchanges of several highly developed economies have
been interpreted using a {\it power-law} probability density function
$P(g) \sim g^{-(\alpha+1)}$ with exponent values $\alpha > 2$, which are
outside the L\'evy-stable regime $0 < \alpha < 2$. To test the
universality of this relationship for less highly developed economies,
we analyze daily returns for the period Nov. 1994---June 2002 for the 49
largest stocks of the National Stock Exchange which has the highest
volume of trade in India. We find that $P(g)$ decays as an {\it
exponential} function $P(g) \sim \exp(-\beta g)$ with a characteristic
decay scales $\beta = 1.51 \pm 0.05$ for the negative tail and $\beta =
1.34 \pm 0.04$ for the positive tail, which is significantly different
from that observed for developed economies. Thus we conclude that the
Indian stock market may belong to a universality class that differs from
those of developed countries analyzed previously.

\end{abstract}


\maketitle


\section{Introduction}

The market index is driven by numerous players and demand-supply factors
through a composite average of various stocks. These factors constitute
the complex market mechanism that causes the price variation in a
component stock, which in turn pulls down or pushes up the a market
index. Tracking many variables is tricky, making the quantification of
economic fluctuations challenging.

A careful analysis of the market forces is required to provide accurate
trends and indicators, which form a tool for market forecast and hence
also provide solutions and key inputs for the improvement of economic
policies and legislation.  In this paper we investigate stock
market asset price variations in a typical developing country such as
India and compare the trends with those from economically developed
economies.


A textbook study~\cite{Hull} of stock price variations suggests that
stock prices---and concomitantly, stock price indices---follow a
Markovian-Wiener process. This means that the stock price on any day is
independent of the history of the stock price or its fluctuation. This
results in a conventional log-normal density for stock
prices~\cite{Hull}, i.e., the logarithm of the stock price follows a
normal density.

However, developed markets such as those in the United States, Germany,
and Japan exhibit a stock price behavior that differs from the Gaussian
density frequently used in conventional theories. A key empirical
finding in this regard is that the probability density of logarithmic
price changes (returns) is approximately symmetric and decays with power
law tails with identical exponent $\alpha \approx 3$ for both
tails~\cite{Lux,Stocks}. One intriguing aspect of this empirical finding
is that it appears to be universal. Individual stocks appear to conform
to these laws not just in US markets~\cite{Stocks}, but also in
German~\cite{Lux} and Australian markets~\cite{Allison}. These same laws
are obeyed by market indices such as the S\&P 500, the Dow Jones, the
NIKKEI, the Hang Seng, and the Milan index~\cite{Index}, and similar
behavior is found in commodity markets~\cite{Kaushik} as well as in the
most-traded currency exchange rates (e.g., the US dollar versus the
Deutsch mark, or the US dollar versus the Japanese yen
~\cite{Dacorogna}). The {\it universal\/} nature of these patterns
exhibited in the statistics of daily returns is remarkable, since these
markets differ greatly in their details. The observed universality is
consistent with a {\it scale-independent\/} behavior of the underlying
dynamics.

\section{Analysis }

Here we focus on Indian stock market and find an exponential probability
density function of price fluctuations, revealing an intrinsic
scale. Our results is based on analyzing $\approx 10^5$ records
representing daily returns for 49 largest stock of the National Stock
Exchange (NSE) in India over the period Nov 1994---June 2002.

We define the normalized price fluctuation (return)
\begin{equation}
g_i(t) \equiv \frac{\log S_i(t+\Delta t)- \log S_i(t)}{\sigma_i}.
\label{e.return}
\end{equation}
Here $\Delta t = 1$ day, $i = 1,2,..,49$ indexes the 49 stocks, $S_i(t)$
is the price of stock $i$ at time $t$, and $\sigma_i $ is the standard
deviation of $\log S_i(t+\Delta t)- \log S_i(t)$. 

\begin{figure}
\narrowtext
\vspace*{0.6cm}
\centerline{
\includegraphics[width=7.5cm,height=7.5cm,angle=-90]{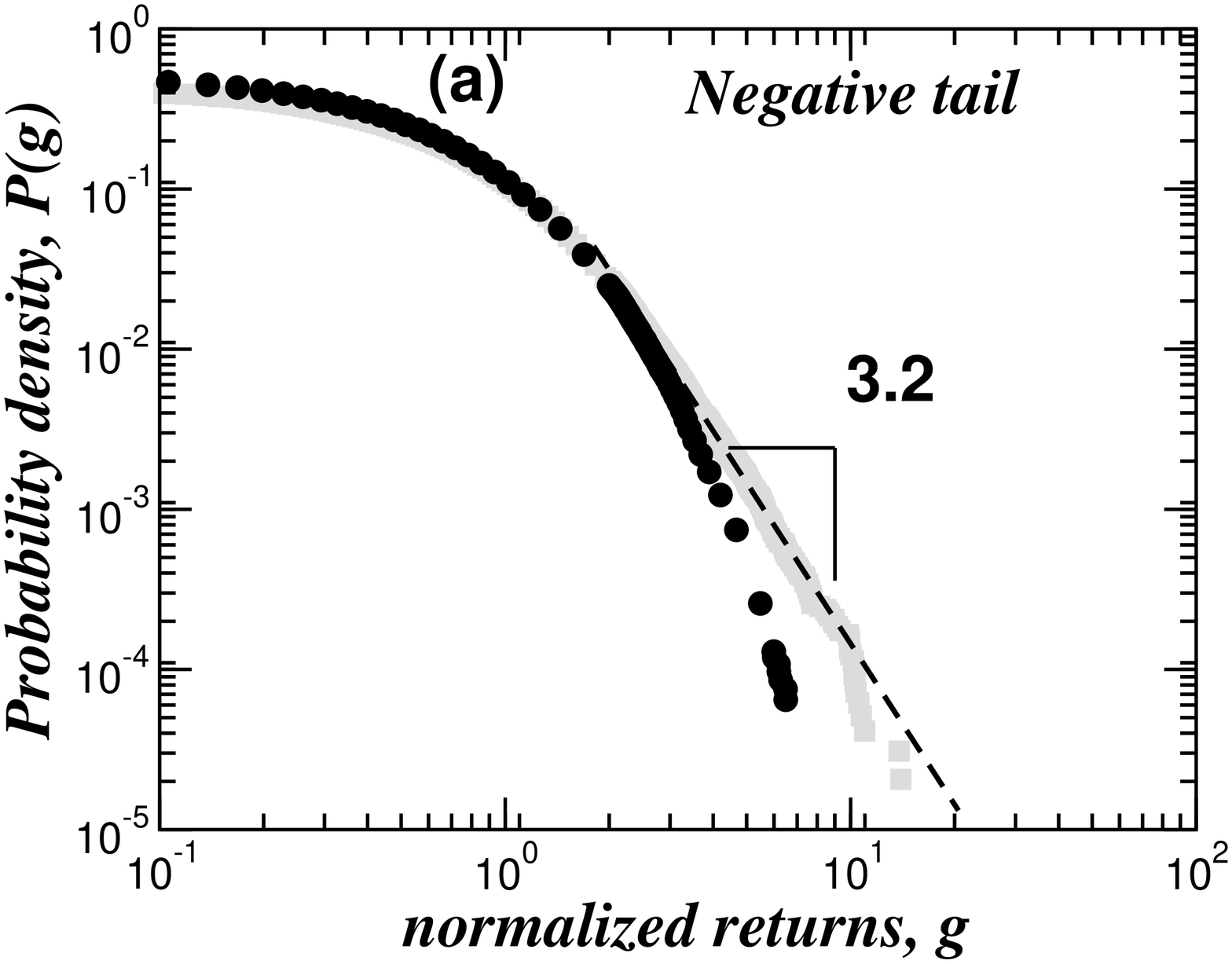}
\includegraphics[width=7.5cm,height=7.5cm,angle=-90]{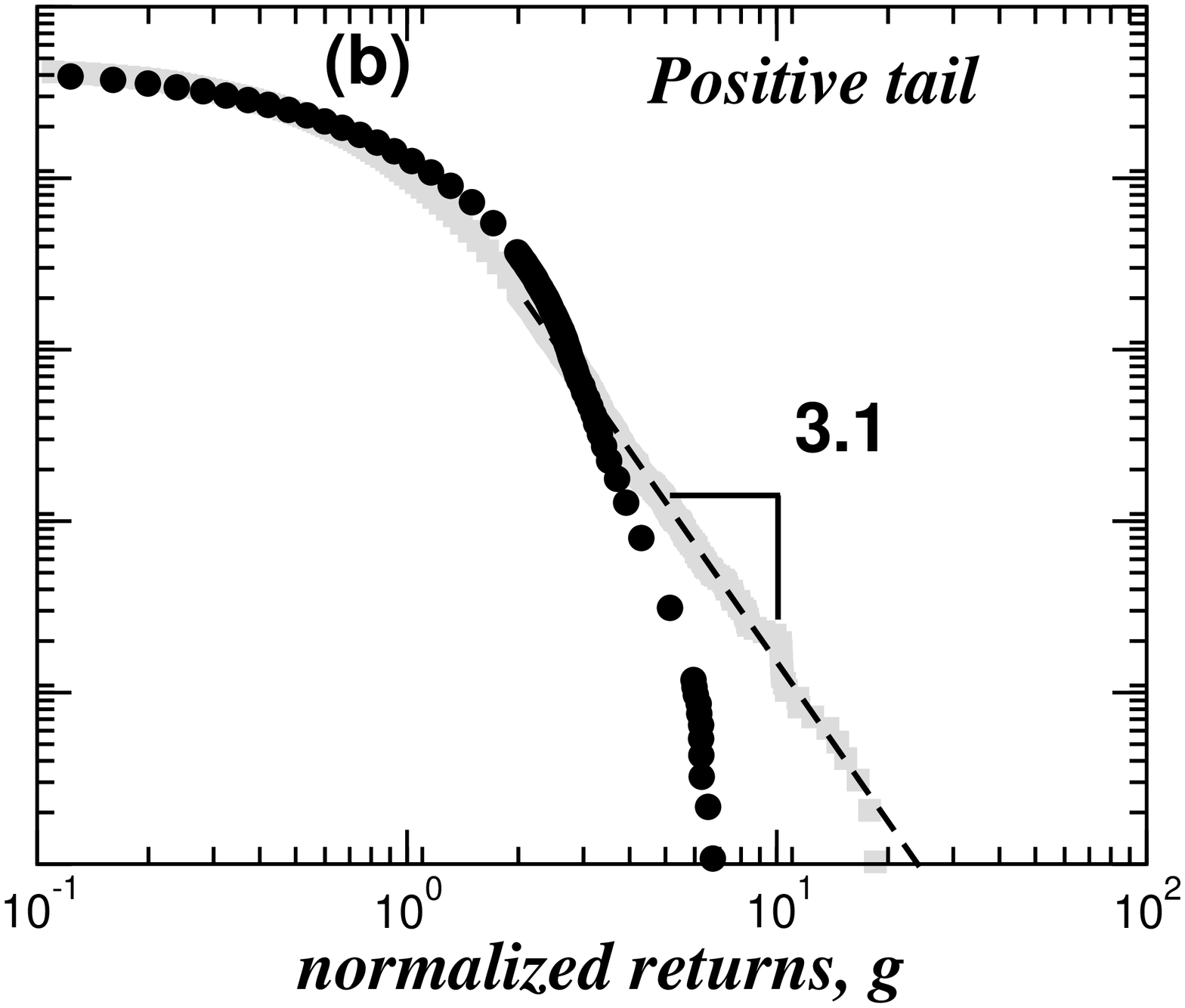}
}
\centerline{
\includegraphics[width=7.5cm,height=7.5cm,angle=-90]{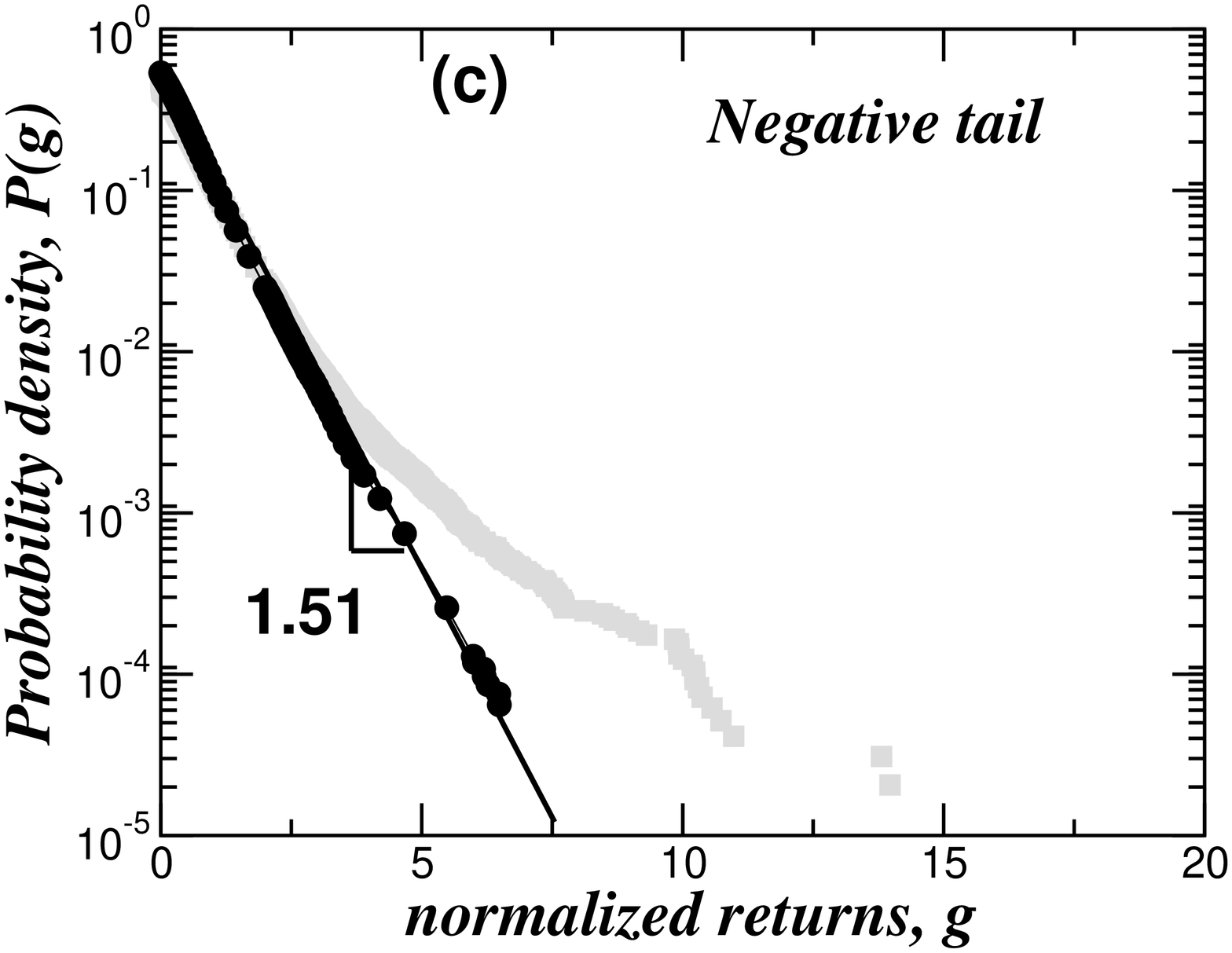}
\includegraphics[width=7.5cm,height=7.5cm,angle=-90]{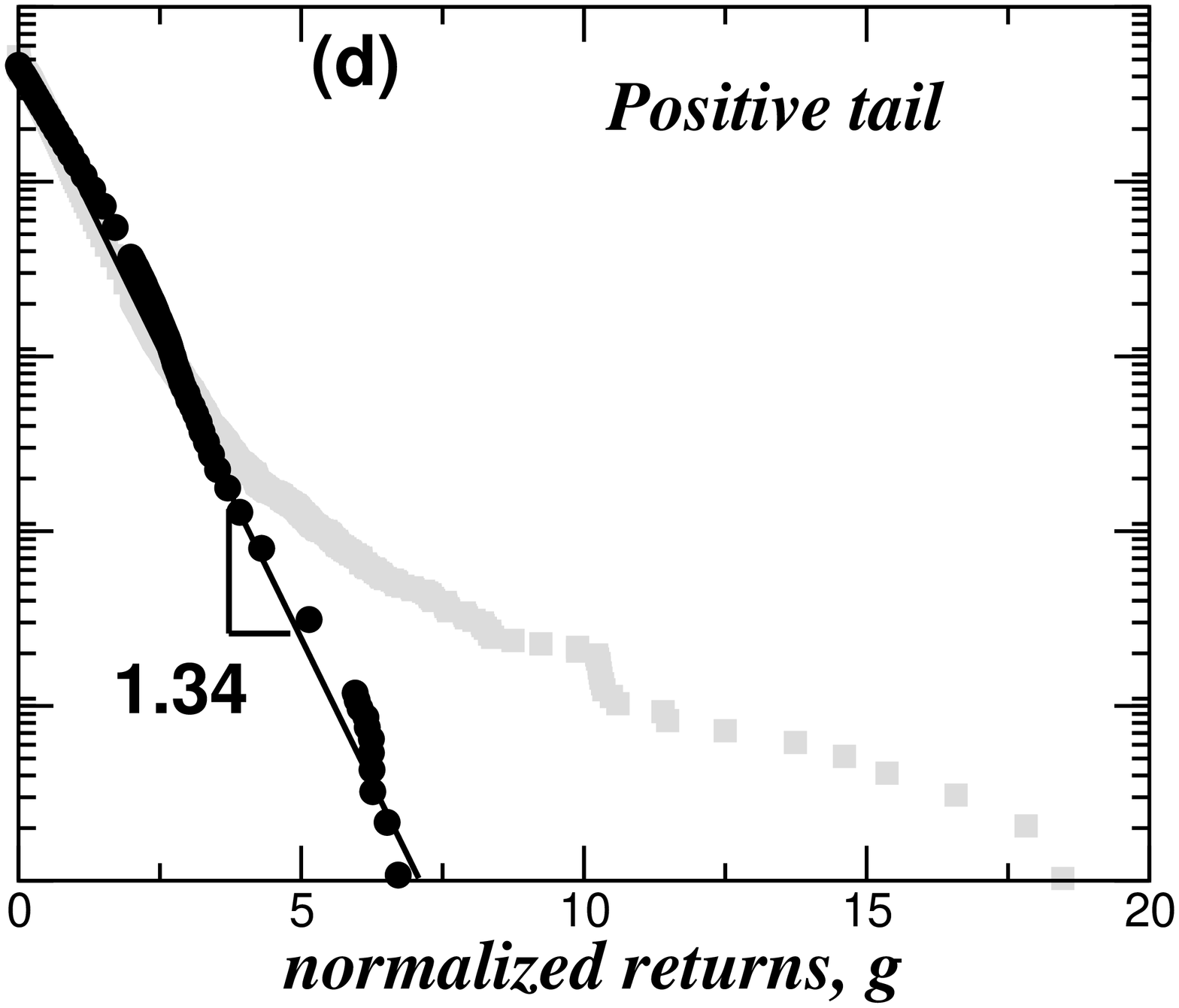}
}
\vspace*{0.1cm}
\caption{ The probability density function of aggregated daily returns
~\protect\cite{note1} on a log-log plot for (a) the negative tail and
(b) the positive tail. Solid symbols are aggregated data from 49 Indian
stocks and the open squares are aggregated data from 49 US stocks over
the same period, Nov. 1994---June 2002. The dashed lines are power law
fits to the US data. The same data for aggregated daily returns on a
linear-log plot for (c) the negative tail and (d) the positive tail. The
solid lines have slopes $\beta = 1.51 \pm 0.05$ for the negative tail
and $\beta = 1.34 \pm 0.04$ for the positive tail where the decay
parameters and the error bars are estimated by the least square
method. The KS significance probability of nullity of the exponential
behavior hypothesis is $ \approx 6.6 \times 10^{-47}$.}
\label{Fig1}
\end{figure}

To compare the probability density function of the Indian stocks with US
stocks we randomly choose 49 US stocks in the same period. Next we
aggregate the data~\cite{note1}. Figures~\ref{Fig1}a and~\ref{Fig1}b
displays the probability density function $P(g)$ for both positive and
negative tails for the daily returns in a log-log plot. The US stocks
have a power law probability density function with exponent $\alpha
\approx 3 $ [cf. ~\cite{Lux,Stocks,Allison,Index}].

Figures~\ref{Fig1}c and~\ref{Fig1}d displays the probability density of
the aggregated data for both Indian and US stocks in a linear-log
plot. We observe that the probability density of the 49 Indian stocks
has an exponential form of decay
\begin{equation}
P(g) \sim e^{-\beta g}
\label{e.pg}
\end{equation}
with 
\begin{equation}
\beta = 
\left\{ \begin{array}{ll}
      1.51 \pm 0.05 & [\mathrm{negative~tail}] \\
      1.34 \pm 0.04 & [\mathrm{positive~tail}] 
       \end{array} \right.
\label{e.beta1}
\end{equation}
 
\begin{figure}
\narrowtext
\vspace*{0.6cm}
\centerline{
\includegraphics[width=9.5cm,height=9.5cm,angle=-90]{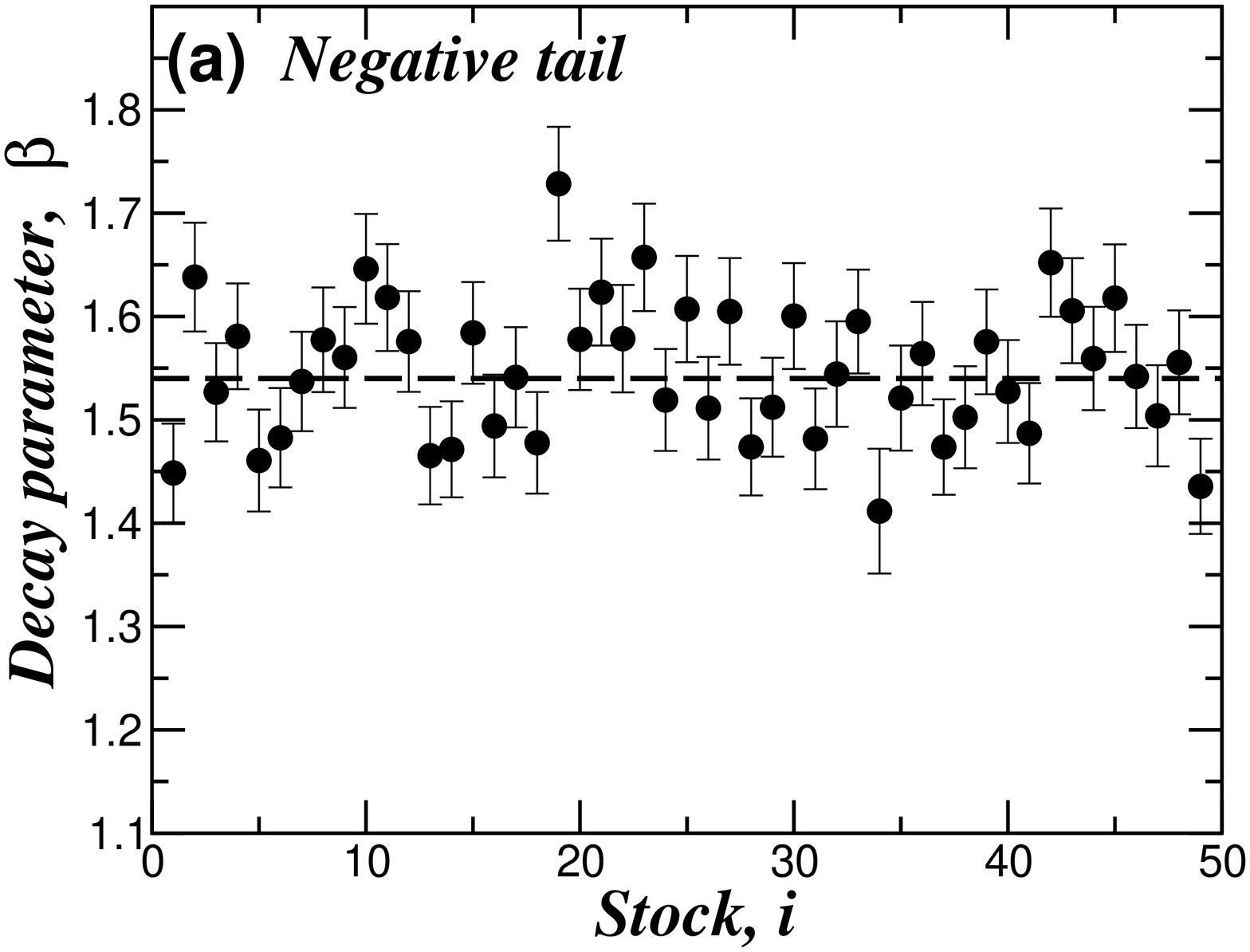}
\includegraphics[width=9.5cm,height=9.5cm,angle=-90]{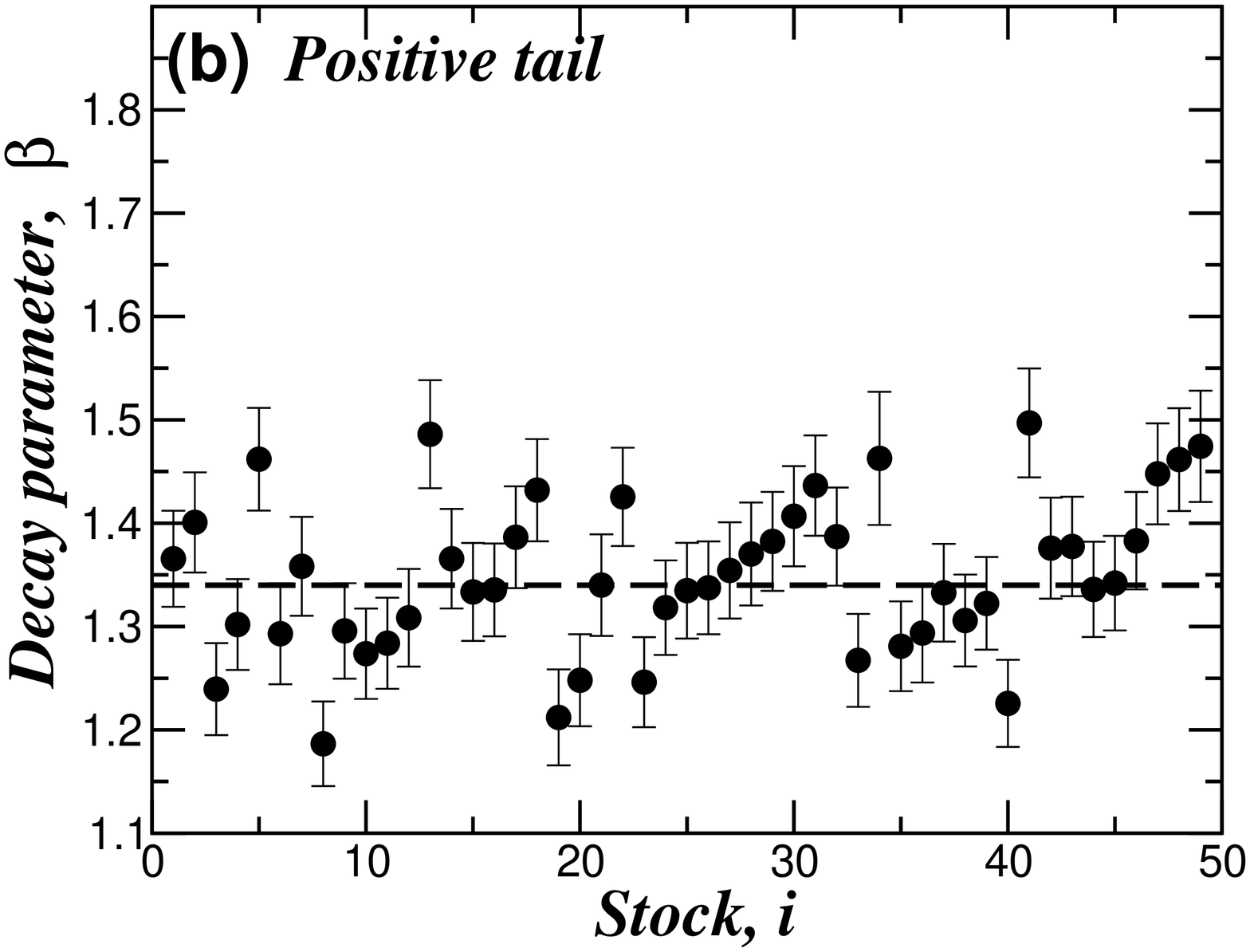}
}
\centerline{
\includegraphics[width=9.5cm,height=9.5cm,angle=-90]{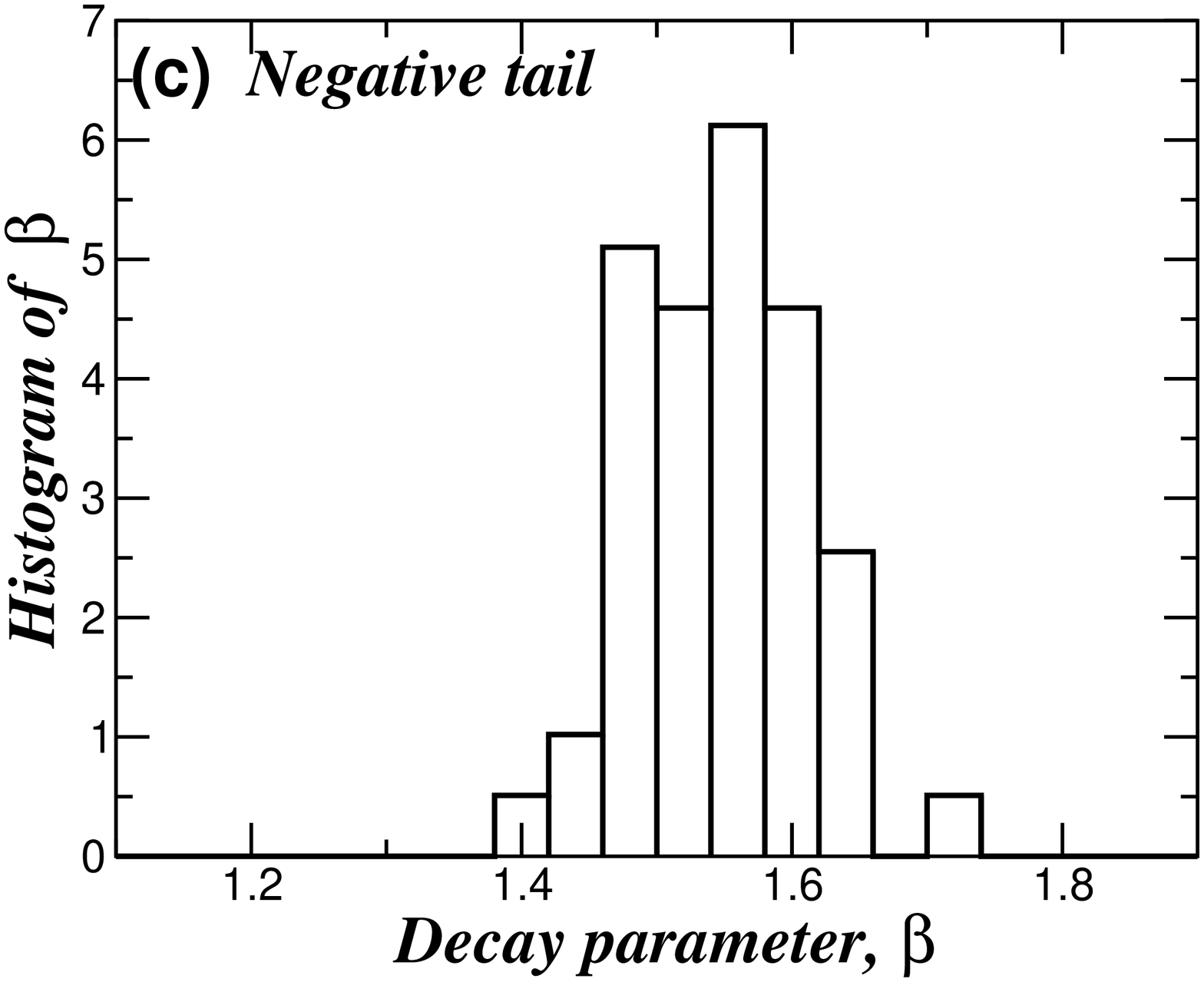}
\includegraphics[width=9.5cm,height=9.5cm,angle=-90]{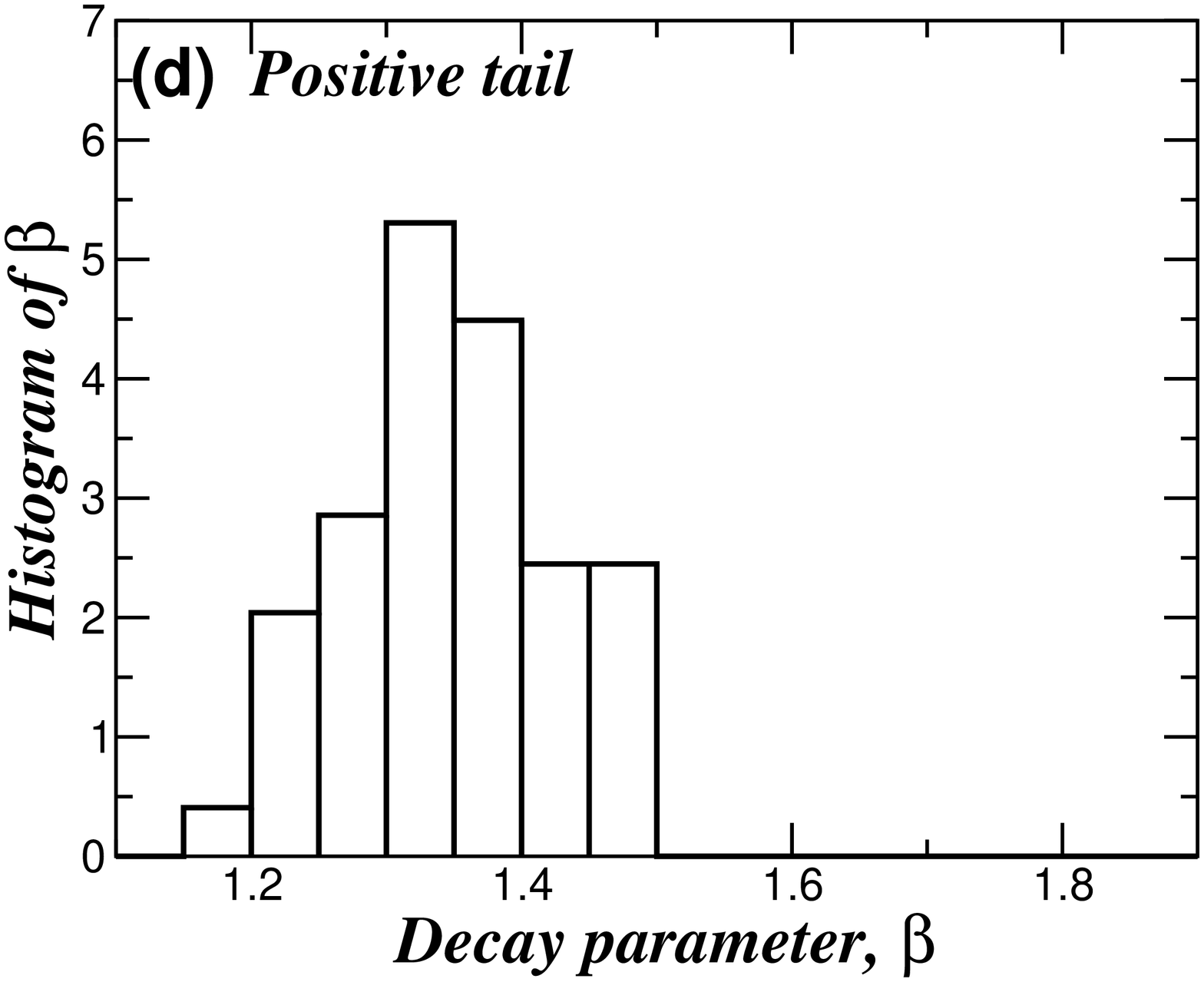}
}
\vspace*{0.1cm}
\caption{ Decay parameters $\beta_i$ of (a) the negative tail and (b)
the positive tail, where $i = 1,2,..,49$ indexes the 49 Indian stocks
analyzed. We employ a least square fit to estimate the parameters
$\beta_i$ of each stock. The dashed lines show the average values
defined in eqs.~\protect\ref{e.beta2}--\protect\ref{e.beta3}. Histogram
of (c) the negative tail decay parameters $\beta_{\rm avg} = 1.54 \pm
0.05$, (d) the positive tail decay parameters$\beta_{\rm avg} = 1.34 \pm
0.05$ .}
\label{Fig2}
\end{figure}

Figure~\ref{Fig2} displays the estimates of $\beta_i$ for both positive
and negative tails of the probability density function. We find the
Kolmogorov-Smirnov (KS) significance probabilities for the null
exponential hypothesis for all 49 Indian stocks and for the aggregated
data to be $\ll 5\%$. Further we calculate
\begin{equation}
\beta_{\rm avg} \equiv \frac{1}{49}\sum_{i=1}^{49} \beta_i 
\label{e.beta2}
\end{equation}
and find
\begin{equation}
\beta_{\rm avg} = 
\left\{ \begin{array}{ll}
      1.54 \pm 0.05 & [\mathrm{negative~tail}] \\
      1.34 \pm 0.06 & [\mathrm{positive~tail}]
       \end{array} \right.
\label{e.beta3}
\end{equation}

\section{Discussion}

Approximately $1/6$ of the world's inhabitants live in India. In 2001
India had an estimated impoverished population of 40 million, 22\% of
the total urban population. The National Stock Exchange averages
$6\times 10^6$ trades per day and its average daily turnover is $\approx
3\times 10^8$ USD. The average turnover in India is $\approx 10^9$ USD
and the average share volume transacted is $\approx 2\times
10^5$. Because Indian people are traditionally extremely careful with
their money, they have a high individual savings and transactions in the
Indian Stock Market are not distributed across all economic
scales. Stock market transactions are typically carried out by those
with wealth in the top 25\% of the economic spectrum.

A natural question is why the Indian stock market should have
statistical properties that differ from other stock markets. One
possible reason can be traced to the history of trading patterns in
India and to its persistent trading culture. Even after more than 127
years of stock market operations, trading in India is said to be based
as much on emotional factors as on actual evaluations and quantitative
analysis.

Quantitative analytical skills, although available, are expensive and
limited, so a majority of investors in India tend to follow archaic
investment strategies, which they feel are more conservative and
safe. The result is that extreme risk situations with concomitant high
returns are completely avoided. This lack of quantification strategies
has also hampered the two year old derivatives market, where even
arbitrageurs trade on thumb rules and not actual models, we have
witnessed prices where mis-pricing takes hours to correct. There are
very few large financial institutions contributing to the total volume
in trade. Small investors drive panic into the market on rumors making
the market susceptible to small instabilities.  Also, until recently,
most Indian assets were under the control of the state and hence exposed
to changes of political administration. These factors have kept the
market under a tight noose.

Thus stock price fluctuations in India are intermediate to that between
power law behavior and Gaussian behavior. Power law behavior is found
for highly developed economies while the less highly developed
economies such as India follow a behavior which is scale
dependent. \footnote{A conjecture would be whether developing economies
which is less developed than India also show Gaussian
behavior~\cite{Hull}. To test this we hope in the future to investigate
whether stock price variations undergo a transition from Gaussian
distribution to a power law distribution via an exponential distribution
at intermediate time}.


We thank L. A. N. Amaral, Y.~Ashkenazy, X.~Gabaix, P.~Gopikrishnan,
S.~Havlin, V.~Plerou, A.~Schweiger and especially T. Lux for helpful
discussions and suggestions, and NSF for financial support. M. P wishes
to acknowledge C. Vasudevan, B. R. Prasad and Manoj Vaish for their kind
encouragement and support.






\begin{references}
\bibitem{note1} We aggregate the data, which is justified if the prices
for all 49 stocks follow the same distribution. This assumption is
consistent with our experience.
\end{references}
\end{document}